\documentclass[journal]{IEEEtran}
\usepackage{amssymb}
\usepackage{amsmath}
\usepackage{bm}
\usepackage{cite}
\usepackage{array}
\usepackage{blkarray}
\usepackage{amsmath}
\usepackage{subfig}
\usepackage{graphicx}
\usepackage{xcolor}
\newcommand{\RN}[1]{%
	\textup{\uppercase\expandafter{\romannumeral#1}}%
}
\ifCLASSINFOpdf
\else
\fi
\begin{document}
	%
	\title{Extreme Learning Machine Based Non-Iterative and Iterative Nonlinearity Mitigation for LED Communications }
	%
	%
	%
	
	\author{Dawei Gao, Qinghua Guo, Jun Tong, Nan Wu, Jiangtao Xi, and Yanguang Yu
		\thanks{D. Gao, Q. Guo, J. Tong, J. Xi and Y. Yu are with the School of Electrical, Computer and Telecommunications Engineering, University of Wollongong, NSW 2522, Australia (e-mail: dg687@uowmail.edu.au; qguo@uow.edu.au; jtong@uow.edu.au; jiangtao@uow.edu.au; yanguang@uow.edu.au).
		}
		\thanks{N. Wu is with the School of Information
			and Electronics, Beijing Institute of Technology, Beijing 100081, China (e-mail: wunan@bit.edu.cn).}}
	%
	%

	\markboth{}r%
	%



	\maketitle
	
	\begin{abstract}
		This work concerns receiver design for light emitting diode (LED) communications where the LED nonlinearity can severely degrade the performance of communications. We propose extreme learning machine (ELM) based non-iterative receivers and iterative receivers to effectively handle the LED nonlinearity and memory effects. For the iterative receiver design, we also develop a data-aided receiver, where data is used as virtual training sequence in ELM training. It is shown that the ELM based receivers significantly outperform conventional polynomial based receivers; iterative receivers can achieve huge performance gain compared to non-iterative receivers; and the data-aided receiver can reduce training overhead considerably. This work can also be extended to radio frequency communications, e.g., to deal with the nonlinearity of power amplifiers.
	\end{abstract}
	

	%
	\IEEEpeerreviewmaketitle

	\section{Introduction}
	%
	%
	%
	%
	\IEEEPARstart{R}{cently}, visible light communication (VLC) has attracted much attention due to the extensive use of light emitting diodes (LEDs) in various lighting applications. VLC is considered a potential complement to the fifth generation (5G) communications as it is viable to alleviate the spectrum scarcity in radio frequency (RF) communications~\cite{6963803}. VLC has been included in the standards such as IEEE 802.15.7, JEITA CP-1221 and JEITA CP-1222, which can be applied in 5G for high data rate transmission~\cite{7593453}. LED is the main light source deployed in VLC and it can provide illumination and data transmission simultaneously due to its fast switching capabilities~\cite{974354}. VLC is considered an energy efficient wireless communication solution and it reduces deployment cost by leveraging existing luminaire infrastructure in 5G green communication systems. LED communications could employ hundreds of terahertz unregulated bandwidth, which is 10,000 times more than the spectrum of RF bands without interfering with the RF bands~\cite{elgala2011indoor}. It is reported that a transmission rate of 3.5 Gb/s was achieved by experiments using a single color incoherent LED~\cite{inproceedings}. In addition, as light is confined to a certain space, employing LEDs is easy to achieve secure transmission and prevent interference from different spaces~\cite{1277847}. 
	
	In LED communications, the light intensity modulation is employed to convert electrical signals to optical signals at a high frequency, which is therefore imperceptible to human eyes. At the receiver side, photo diodes (PDs) are used to detect the light intensity and convert it back to electrical signals. However, LED exhibits nonlinearity, which can significantly degrade the communication performance, in particular, when high order modulations such as pulse amplitude modulation (PAM) are employed to achieve high speed transmission~\cite{5259919,7096283}. The nonlinear behavior of LED imposes amplitude distortion on the input signals. The lower peaks of signals are clipped at the LED turn-on voltage (TOV), and the upper peaks are saturated at the maximum allowable voltage recommended by the manufacturer~\cite{7096283}.
	
	A variety of nonlinearity mitigation techniques have been developed in the literature, which can be divided into two categories: pre-distortion at transmitter and post-distortion at receiver. The pre-distortion techniques require additional feedback link from receiver to transmitter or assume that the LED nonlinearity is known and fixed. The LED nonlinearity without memory effects was considered in~\cite{5259919} and~\cite{elgala2009non}, where a polynomial-based pre-distorter was proposed to compensate the LED nonlinearity. However, LED exhibits not only instantaneous nonlinearity, but also memory effect in high speed communications. A Volterra series based equalizer was proposed in~\cite{5771519} to mitigate LED nonlinearity with memory effects. However, the determination of Volterra coefficients requires high computational cost. To circumvent this issue, a memory polynomial based technique was proposed to post-distort the LED nonlinearity at the receiver~\cite{qian2014adaptive}. However, these polynomial based methods involve the determination of polynomial coefficients with a matrix inverse operation, which can easily lead to numerical instability due to the ill condition of the matrix~\cite{1337325,2002194,1337247}. Meanwhile, as a nonlinear system may not be thoroughly compensated by a finite order polynomial, errors are inevitably incurred~\cite{20546}. As post-distortion is deployed at the receiver side, it is more convenient to implement and is more suitable to handle the time-varying LED nonlinearity and VLC channel. Hence, we focus on post-distortion in this paper. 
	
	In this paper, we first propose an extreme learning machine (ELM) based non-iterative receiver to jointly mitigate LED nonlinearity and memory effects due to LED and VLC channel. Then, we consider iterative nonlinearity mitigation and decoding in a coded LED communication system. We design iterative receivers, which consists of an ELM based soft-in-soft-out (SISO) post-distorter and a SISO decoder. The SISO post-distorter and the decoder work in an iterative manner by exchanging log-likelihood ratios (LLRs) of coded bits. The ELM based post-distorter computes the extrinsic LLRs of coded bits based on the feedback from the SISO decoder and observations where ELM is used to handle both LED nonlinearity and memory effects. The ELM in the post-distorter can be trained with training sequence. To improve the transmission efficiency by reducing the training overhead, we further investigate a data-aided ELM based post-distorter, where both training sequence and data are utilized to train the ELM. We show that the training overhead can be significantly reduced in the data-aided ELM based scheme. It is shown that compared with the conventional polynomial based techniques, ELM is much more effective to combat the LED nonlinearity, which is demonstrated in Section \RN{5}. It is noted that this work can be extended to RF communications, e.g., to handle the nonlinearity of power amplifiers. 
	
	The rest of the paper is organized as follows. Section \RN{2} gives a brief review of the ELM. Section \RN{3} elaborates the signal model and details the design of ELM based non-iterative and iterative receivers for LED nonlinearity mitigation and decoding. In Section \RN{4}, a data-aided ELM based iterative receiver is designed and discussed. In Section \RN{5}, experimental results are provided to verify the effectiveness of the proposed receivers. Conclusions are drawn in Section \RN{6}.
	
	\section{Extreme Learning Machine}
	\begin{figure}
		\begin{center}
			\includegraphics[width=3.0in]{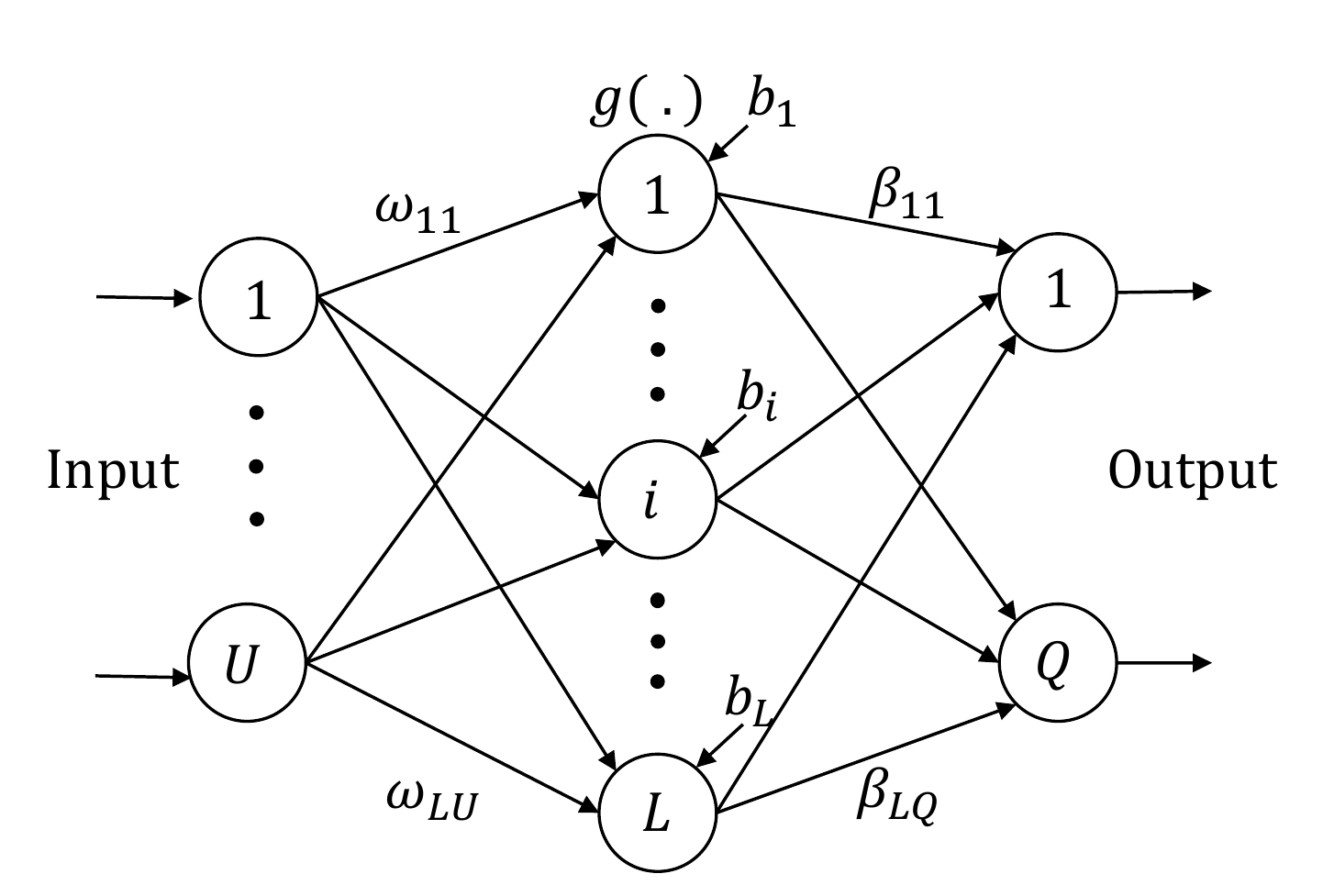}\\
			\caption{Structure of ELM.}\label{ELM}
		\end{center}
	\end{figure}
	As shown in Fig.~\ref{ELM}, ELM is a single-hidden layer feedforward neural network, where the hidden nodes (i.e., the input weights $\{\omega_{ij}\}$ and biases $\{b_i\}$) are randomly initialized and fixed without tuning~\cite{huang2006extreme}. The parameters to be learned in ELM are the output weights, and hence ELM can be formulated as a linear model with respect to the parameters, which boils down to solving a linear system, making ELM efficient in learning.
	For $N$ distinct training samples $\{(\bm{s}_j,\bm{y}_j)\in\mathbb{R}^U\times \mathbb{R}^Q \}_{j=1}^N$, where $\bm{s}_j=[s_{j1},s_{j2},\ldots,s_{j(U-1)},s_{jU}]^T$ and $\bm{y}_j=[{y}_{j1},{y}_{j2},\ldots,{y}_{j(Q-1)},{y}_{jQ}]^T$, the output of the SLFN shown in Fig.~\ref{ELM} can be expressed as
	\begin{equation}\label{es}
	\bm{\psi}_j = \sum_{i=1}^{{L}} \bm{\beta}_i g(\bm{\omega}_i^T\bm{s}_j+b_i), \hskip 1pc j = 1,\dotsc,N,
	\end{equation}
	where $L$ is the number of hidden nodes, $\bm{\omega_i}=[\omega_{i1},\omega_{i2},\ldots,\omega_{iU}]^T$ is the input weight vector that connects all input nodes to the $i$th hidden node, $b_i$ is the bias of the $i$th hidden node, and $g(.)$ is the activation function of the hidden layer. $\bm{\beta}_i=[{\beta}_{i1},{\beta}_{i2},\ldots,{\beta}_{iQ}]^T$ denotes the output weight vector connecting the $i$th hidden node and the output nodes. The output nodes are chosen to be linear in this network. 
	
	By incorporating the $N$ equations in~(\ref{es}), it can be represented compactly as 
	\begin{equation}\label{esxme}
	\bm{H}\bm{\beta}=\bm{\Psi},
	\end{equation}
	where
	\begin{equation}
	\bm{H}=
	\begin{bmatrix}
	g(\bm{\omega}_1^T\bm{s}_1+b_1) & \cdots & g(\bm{\omega}_L^T\bm{s}_1+b_L)  \\
	\vdots & \ddots & \vdots \\
	g(\bm{\omega}_1^T\bm{s}_N+b_1) & \cdots & g(\bm{\omega}_L^T\bm{s}_N+b_L)  \end{bmatrix}_{N\times L},
	\end{equation}
	\[
	\bm{\beta}=
	\begin{bmatrix}
	\bm{\beta}_1^T \\
	\vdots\\
	\bm{\beta}_L^T
	\end{bmatrix}_{L \times Q},\quad
	\bm{\Psi}=
	\begin{bmatrix}
	\bm{\psi_}1^T  \\
	\vdots\\
	\bm{\psi}_N^T
	\end{bmatrix}_{N \times Q},
	\]
	and $\bm{H}$ denotes the hidden layer output matrix.
	
	ELM randomly selects input weights and biases, and output weights $\bm{\beta}$ are obtained by minimize the cost function $\sum_{j=1}^N  \left\lVert \bm{\psi}_j-\bm{y}_j \right\rVert^2$, which is given by the least-squares (LS) solution, i.e., 
	\begin{equation}\label{beta}
	\bm{\beta}=\bm{H}^\dagger \bm{Y},
	\end{equation}
	where $\bm{H}^\dagger$ is the Moore-Penrose generalized inverse of matrix $\bm{H}$ and $\bm{Y} =[\bm{y}^T_1,\bm{y}^T_1,\ldots,\bm{y}^T_N]^T$.

	\section{ELM-Based Non-iterative and Iterative Receiver Design}
	\subsection{Signal Model}
	We consider a single-carrier coded LED communication system with PAM, which is shown in Fig.~\ref{Block_pre} for a system with a non-iterative receiver and Fig.~\ref{Block} with an iterative receiver. At the transmitter side, the information bits are first encoded into a code sequence (and permuted by an interleaver in an iterative LED system as shown in Fig.~\ref{Block}). The encoded (or interleaved) code sequence $\bm{c}$ is split into a number of length-$P$ subsequences $\{\bm{c}_n=[c_{n,1},c_{n,2},\ldots,c_{n,P}]\}$, and each subsequence $\bm{c}_n$ is mapped to a symbol ${x}_n \in D$ with a mapping rule (e.g., Gray mapping), where $D=\{\alpha_i,i=1,2,\ldots,2^P \}$ denotes a $2^P$-ary PAM symbol alphabet. Symbol ${x}_n$ is used to modulate the LED light intensity for transmission. As a result of the LED nonlinearity and memory effect, the transmitted signal is expressed as
	\begin{equation}
	z_n=f({x}_n,{x}_{n-1},\ldots,{x}_{n-M}),
	\end{equation}
	where $f(.)$ represents the LED nonlinear distortion with memory effect and $M$ is the memory length. The memory polynomial model is widely used in the literature, and in this case the LED nonlinearity with memory effect can be modelled as~\cite{7096283}
	\begin{equation}\label{MP}
	z_n=\sum_{k=1}^K\sum_{m=0}^Ma_{k,m} {x}_{n-m}^k, 
	\end{equation}
	where $K$ is the order of the memory polynomial, and $\{a_{k,m}\}$ are the coefficients of the polynomial. It is noted that the memory effect of VLC channel, which causes inter-symbol interference (ISI), can be absorbed into the memory polynomial model. The signal is captured by the PD and converted to an electrical signal. We assume a unit PD responsivity for simplicity, and the received signal can be represented as
	\begin{equation}\label{PN}
	{y}_n=z_n+w_n,
	\end{equation}
	where $w_n$ is the additive white Gaussian noise (AWGN) with zero mean and variance $\sigma^2$.
	\subsection{ELM Based Non-iterative Receiver}\label{pol}
	\begin{figure}
		\begin{center}
			\includegraphics[width=3.5in]{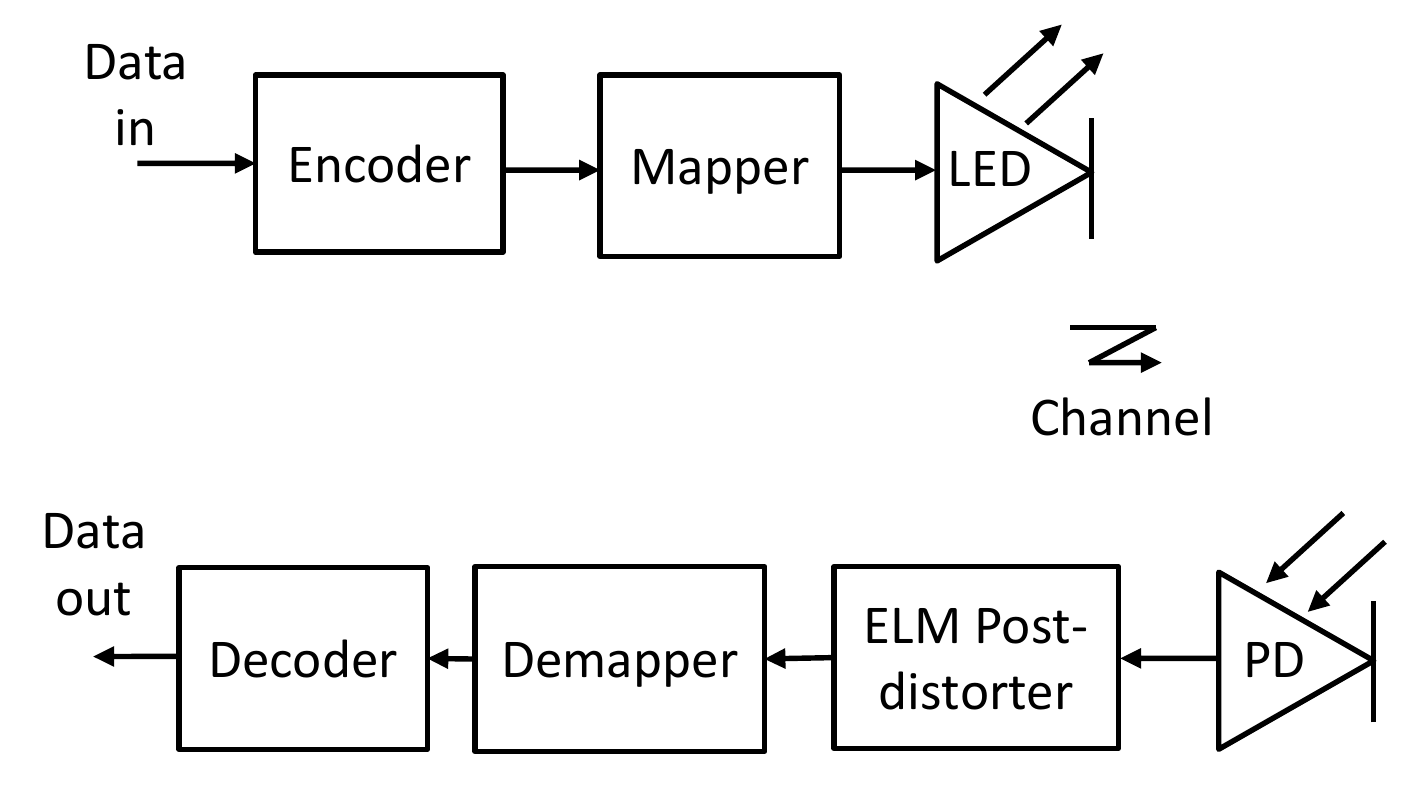}\\
			\caption{Block diagram of a coded LED communication system with an ELM based non-iterative receiver.}\label{Block_pre}
		\end{center}
	\end{figure}
	
	We first introduce the conventional polynomial based non-iterative receiver.  For an LED system with nonlinearity and memory effects, memory polynomial based post-distorter has been investigated~\cite{qian2014adaptive}. With the received signal $\{y_n\}$ as input, the output of the post-distorter can be represented as~\cite{qian2014adaptive},
	\begin{equation}\label{memp}
	\hat{x}_n=\sum_{k=1}^{\Tilde{K}}\sum_{m=0}^{\Tilde{M}}a^{\ast}_{k,m} {y}_{n-m}^k,  \hskip 1pc n = 1,\dotsc,N,
	\end{equation}
	where $\Tilde{K}$ is the order of the memory polynomial used in the post-distorter, $\Tilde{M}$ is the memory length, and $\{a^{\ast}_{k,m}\}$ are the coefficients of the polynomial. We can rewrite (\ref{memp}) in a vector form as
	\begin{equation}\label{mempstar}
	\hat{x}_n=\bm{r}_n^T\bm{a}^{\ast},
	\end{equation}
	where 
	\begin{equation}\label{a_star}
	\bm{a}^{\ast} = [{a}^{\ast}_{1,0},{a}^{\ast}_{2,0},\ldots,{a}^{\ast}_{\Tilde{K},0},\ldots,{a}^{\ast}_{1,\Tilde{M}},{a}^{\ast}_{2,\Tilde{M}},\ldots,{a}^{\ast}_{\Tilde{K},\Tilde{M}}]^T,
	\end{equation}
	and
	\begin{equation}\label{y_vn}
	\bm{r}_n = [y_n,y^2_n,\ldots,y^{\Tilde{K}}_n,\ldots,y_{n-\Tilde{M}},y^2_{n-\Tilde{M}},\ldots,y^{\Tilde{K}}_{n-\Tilde{M}}]^T.
	\end{equation}
	By minimizing the cost function $\sum_{n=1}^N  \left\lVert {x}_n-\bm{r}^T\bm{a}^{\ast} \right\rVert^2$, the LS solution to the coefficient vector $\bm{a}^{\ast}$ can be calculated as~\cite{qian2014adaptive},
	
	\begin{equation}\label{as}
	\bm{a}^{\ast} = (\bm{R}^T\bm{R})^{-1}\bm{R}^T\bm{x}'
	\end{equation}
	where 
	\begin{equation}\label{xprime}
	\bm{x}' = [x'_1, x'_2,\ldots,x'_N]^T
	\end{equation}
	is the training sequence, and 
	\begin{equation}\label{yv}
	\bm{R} = [\bm{r}_1^T, \bm{r}_2^T,\ldots,\bm{r}_N^T]^T
	\end{equation} 
	is constructed based on the received sequence $\{y'_n\}$.
	The coefficient vector $\bm{a}^{\ast}$ can also be calculated recursively by using the recursive least squares (RLS)~\cite{qian2014adaptive}.  
	
\begin{figure}
		\begin{center}
			\includegraphics[width=3.0in]{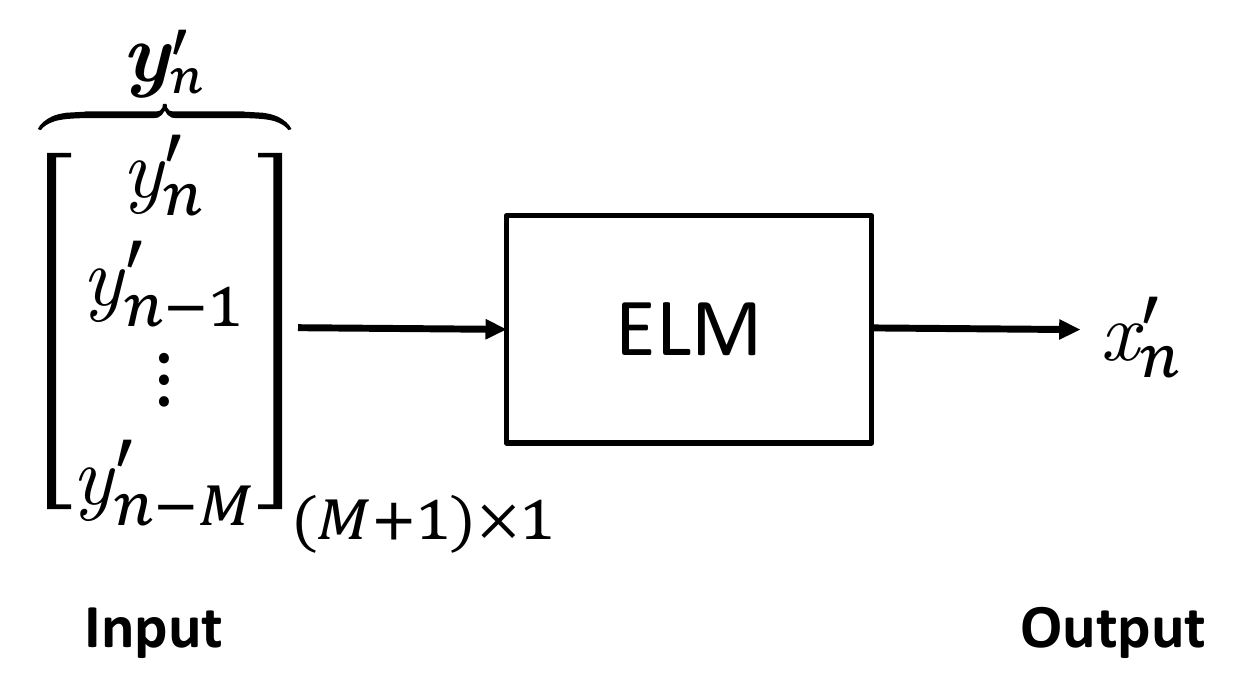}\\
			\caption{Training of ELM based non-iterative post-distorter where ELM has a single output.}\label{ELMni}
		\end{center}
	\end{figure}

	It has been shown that the matrix inverse in~(\ref{as}) suffers from numerical instability as $\bm{R}^T\bm{R}$ is usually an ill-conditioned matrix~\cite{1337325}. It is also noted that to enable the polynomial based post-distorter to have a high capability of LED nonlinearity compensation, a high order polynomial is required. However, a higher order polynomial can more easily lead to ill condition of the matrix $\bm{R}^T\bm{R}$. This can result in significant performance degradation~\cite{2002194,1337247}. In this work, we first propose an ELM based non-iterative receiver for the mitigation of LED nonlinearity and memory effects. The system diagram is shown in Fig.~\ref{Block_pre}, where ELM works as a post-distorter to compensate the LED nonlinearity and memory effects. Here, we use a sliding window approach, i.e., the input of the ELM is a vector $\bm{y}^{\prime}_n= [{y}^{\prime}_n,{y}^{\prime}_{n-1},\ldots,{y}^{\prime}_{n-M}]^T$, which is obtained by windowing the received signal sequence, and the expected output of ELM is a single symbol ${x}^{\prime}_n$ in training. This is shown in Fig.~\ref{ELMni}, where the input and output dimensions of ELM are $M+1$ and $1$, respectively. Once the ELM post-distorter is trained using~(\ref{beta}) and the trained output weights $\bm{\beta}^{\ast}=[\beta^{\ast}_1,\beta^{\ast}_2,\ldots,\beta^{\ast}_L]^T$ is obtained, it can be used to predict the transmitted symbols $\{{x}_n\}$ based on the received signal $\{{y}_n\}$, i.e.,
	\begin{equation}\label{esss}
	{\hat{x}}_n= \sum_{i=1}^{{L}} \beta_i^\ast g(\bm{\omega}_i^T{\bm{y}}_n+b_i), \hskip 1pc n = 1,\dotsc,N,
	\end{equation}
	where $\bm{y}_n = [{y}_n,{y}_{n-1},\ldots,{y}_{n-M}]^T$.
	
	It turns out that ELM based post-distorter is much more powerful than the conventional memory polynomial based post-distorter in dealing with LED nonlinearity. The computation of the memory polynomial coefficients can easily suffer from numerical instability even for moderate order of the memory polynomial, which can severely impact the performance of memory polynomial based post-distorter. In contrast, the ELM based post-distorter works very well as demonstrated in Section \RN{5}.  
	\subsection{ELM Based Iterative Receiver}\label{ss}
\begin{figure}
		\begin{center}
			\includegraphics[width=3.5in]{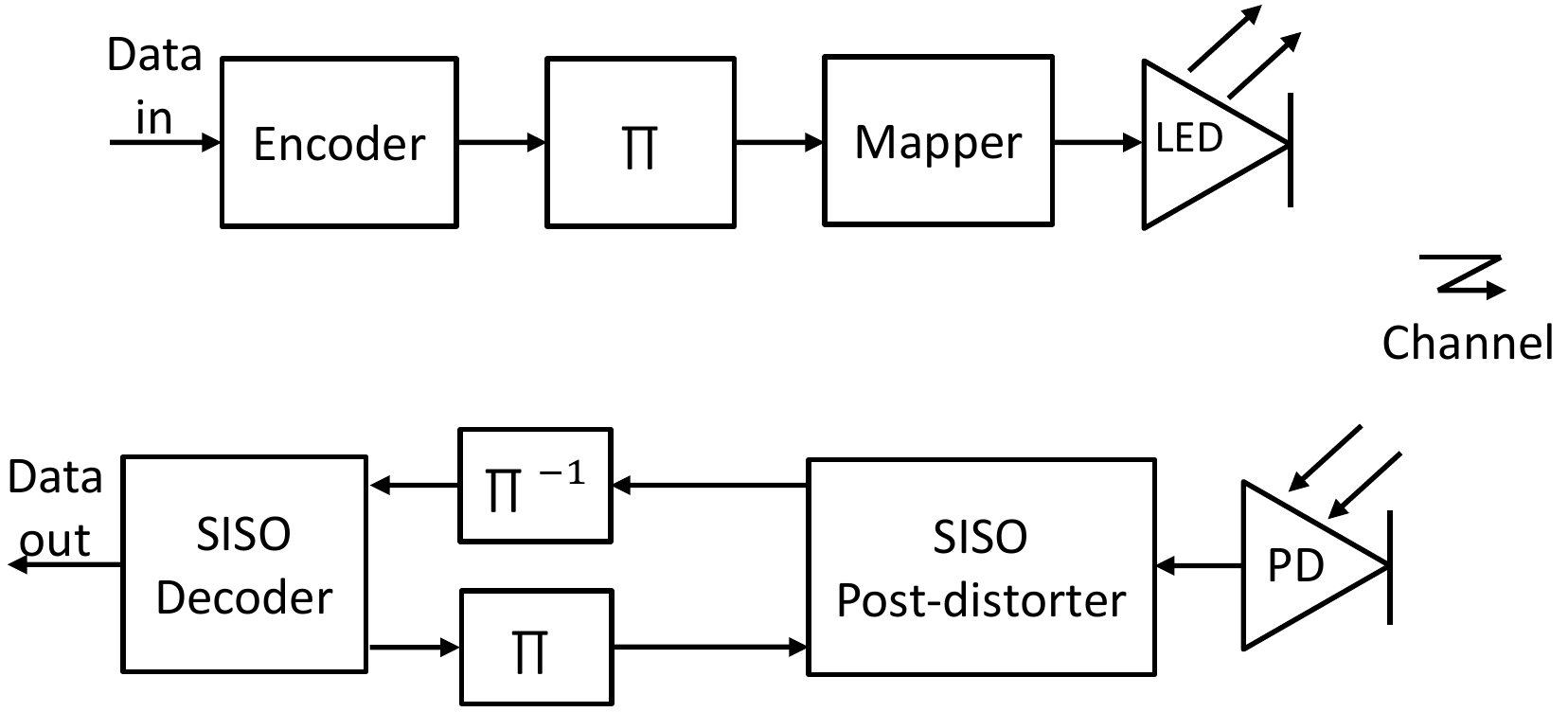}\\
			\caption{Block diagram of iterative nonlinearity mitigation and decoding, where $\Pi$ and $\Pi^{-1}$  denote an interleaver and the corresponding deinterleaver, respectively.}\label{Block}
		\end{center}
	\end{figure}

	It is well known that iterative receivers are powerful to combat ISI in wireless communications~\cite{060506,5741768,1006557}. Inspired by this, we also propose an ELM based iterative receiver for the mitigation of LED nonlinearity and memory effects. The iterative receiver is shown in the lower part of Fig.~\ref{Block}, which is composed of a SISO post-distorter and a SISO decoder, and they work in an iterative manner by exchanging the extrinsic LLRs of coded bits. The SISO post-distorter calculates the extrinsic LLRs for each coded bit with the extrinsic LLRs from the decoder as the \textit{a priori} information. Then, with the extrinsic LLRs from the post-distorter, the decoder refines the LLRs with the code constraints.
	In this work, standard SISO decoding algorithms (e.g., the BCJR algorithm for convolutional codes) are employed and we focus on the design of the SISO post-distorter. The aim of the post-distorter is to calculate the LLR for each code bit $c_{n,p}$, which can be represented as
	\begin{align}
	{LLR}^p(c_{n,p}) &  = \ln{\frac{p(c_{n,p}=0\lvert\bm{y})}{p(c_{n,p}=1\lvert\bm{y})}}\\        
	&	=\ln{\frac{\sum_{\alpha_i\in D_p^0}p({x}_n=\alpha_i\lvert\bm{y})}{\sum_{\alpha_i\in D_p^1}p({x}_n=\alpha_i\lvert\bm{y})}},\label{LLR}
	\end{align}
	where $D_p^0$ and $D_p^1$ denote the subsets of all $\alpha_i\in D$ whose label in position $p$ has the value of 0 and 1, respectively. $\bm{y}=[{y}_1,{y}_2,\ldots,{y}_N]$ is the received signal vector. According to the turbo principle~\cite{1006557}, extrinsic LLR for each coded bit is passed to the decoder, which can be expressed as 
	\begin{equation}\label{LLR2}
	{LLR}^e(c_{n,p})= {LLR}^p(c_{n,p})-{LLR}^a(c_{n,p}),
	\end{equation}
	where ${LLR}^a(c_{n,p})$ is the \textit{a priori} LLR from the decoder in the previous iteration.
	
	It can be found from~(\ref{LLR}) that the key of the post-distorter is to compute the \textit{a posteriori} probability of each transmitted symbol $p(x_n=\alpha_i\lvert \bm{y})$. Based on the Bayes’ theorem, we have 
	\begin{equation}
	p(x_n=\alpha_i\lvert \bm{y})\propto p(\bm{y}\lvert x_n )p(x_n),
	\end{equation}
	where $p(x_n )$ is the \textit{a priori} probability of symbol $x_n$, and it can be computed based on the feedback from the decoder, i.e.,
	\begin{equation}
	p(x_n=\alpha_i)=\prod_{j=1}^{L}p(c_{n,j}=\Omega_{i,j}),
	\end{equation}
	where each $\alpha_i$ corresponds to a binary vector $\bm{\Omega}_i=[\Omega_{i,1},\Omega_{i,2},\ldots,\Omega_{i,L}]$. Next, we need to compute the likelihood $p(\bm{y}\lvert x_n )$ for each $x_n$. 
	
	We use a sliding window approach to approximate the likelihood. Define ${\bm{y}}_n=[{y}_n,{y}_{n+1},\ldots,{y}_{n+M}]^T$ and $\bm{w}_n=[{w}_n,{w}_{n+1},\ldots,{w}_{n+M}]^T$. We compute $p(\bm{y}_n\lvert x_n )$ instead of $p(\bm{y}\lvert x_n )$. Based on~(\ref{MP}) and~(\ref{PN}), we have
	
	\begin{equation}\label{MI}
	\bm{y}_n=
	\underbrace{\begin{bmatrix}
		\sum\limits_{i=1}^Ka_{k,0}x_{n}^k \\
		\vdots \\
		\sum\limits_{k=1}^Ka_{k,q}x_{n}^k \\
		\vdots \\
		\sum\limits_{k=1}^Ka_{k,M}x_{n}^k  
		\end{bmatrix}}_{\bm{a}({x_n})}
	+
	\underbrace{\begin{bmatrix}
		\sum\limits_{k=1}^K\sum\limits_{m=1}^Ma_{k,m}x_{n-m}^k     \\
		\vdots \\
		\sum\limits_{k=1}^K\sum\limits_{m=0,m\neq q}^Ma_{k,m}x_{n+q-m}^k    \\
		\vdots \\  
		\sum\limits_{k=1}^K\sum\limits_{m=0}^{M-1}a_{k,m}x_{n+q-m}^k
		\end{bmatrix}}_{\mathbf{{i}}_{n}}
	+\bm{w}_n,
	\end{equation}
	where $0<q<M$, $\bm{a}(x_n)$ is the useful signal related to $x_n$, and $\mathbf{i}_{n}$ is the interference with respect to $x_n$.
	
	We approximate the interference $\mathbf{{i}}_{n}$ to be Gaussian with mean vector  $\overline{\mathbf{{i}}}_{n}$ and covariance matrix $\bm{V}_{\bm{\mathrm{i}}_n}$. Then, based on~(\ref{MI}), we have
	\begin{equation}\label{pyx}
	p(\bm{y}_n\lvert x_n )\propto \exp(-\frac{1}{2}[\bm{y}_n-\bm{m}]^T\bm{V}^{-1}[\bm{y}_n-\bm{m}]),
	\end{equation}
	where
	\begin{equation}\label{m}
	\bm{m} = \bm{a}(\alpha_i)+\overline{\mathbf{{i}}}_{n},
	\end{equation}
	and $\bm{V}$ is the covariance matrix of interference $\mathbf{{i}}_{n}$ plus noise $\bm{w}_n$ in~(\ref{PN}), i.e.,
	\begin{equation}\label{V}
	\bm{V} =\bm{V}_{\bm{\mathrm{i}}_n} +\sigma^2\bm{I}.
	\end{equation}
	
	To obtain the likelihood in~(\ref{pyx}), we need to compute $\bm{a}(\alpha_i)$, $\overline{\mathbf{{i}}}_{n}$ and $\bm{V}$. It can be seen that all of them depend on the LED nonlinearity, i.e., the memory polynomial coefficients $\{a_{k,m}\}$. In~\cite{8451951}, the iterative receiver is designed with the perfect knowledge of of LED nonlinearity and memory effects, i.e., assuming that the memory polynomial coefficients $\{a_{k,m}\}$ are perfectly known. However, they are usually unknown in practice. A possible solution is to first estimate the polynomial coefficients $\{a_{k,m}\}$ in~(\ref{MP}) and then calculate $\bm{a}(\alpha_i)$, $\overline{\mathbf{{i}}}_{n}$ and $\bm{V}$. However, it can be difficult since the estimation of memory polynomial coefficients involves the inversion of a correlation matrix with ill condition, which can easily lead to numerical instabilities especially for high order or even moderate order polynomials~\cite{zhao2016orthogonal,2002194}. In addition, it can also be difficult to determine the polynomial order and memory length to properly model the LED nonlinearity and memory effects.
	
	In this work, we use ELM to efficiently model the LED nonlinearity. The ELM is trained with a training sequence, and then $\bm{a}(\alpha_i)$, $\overline{\mathbf{{i}}}_{n}$ and $\bm{V}$ can be estimated based on the trained ELM, which are detailed in the following. 
	
	\subsubsection{ELM training} 
\begin{figure}
		\begin{center}
			\includegraphics[width=3.0in]{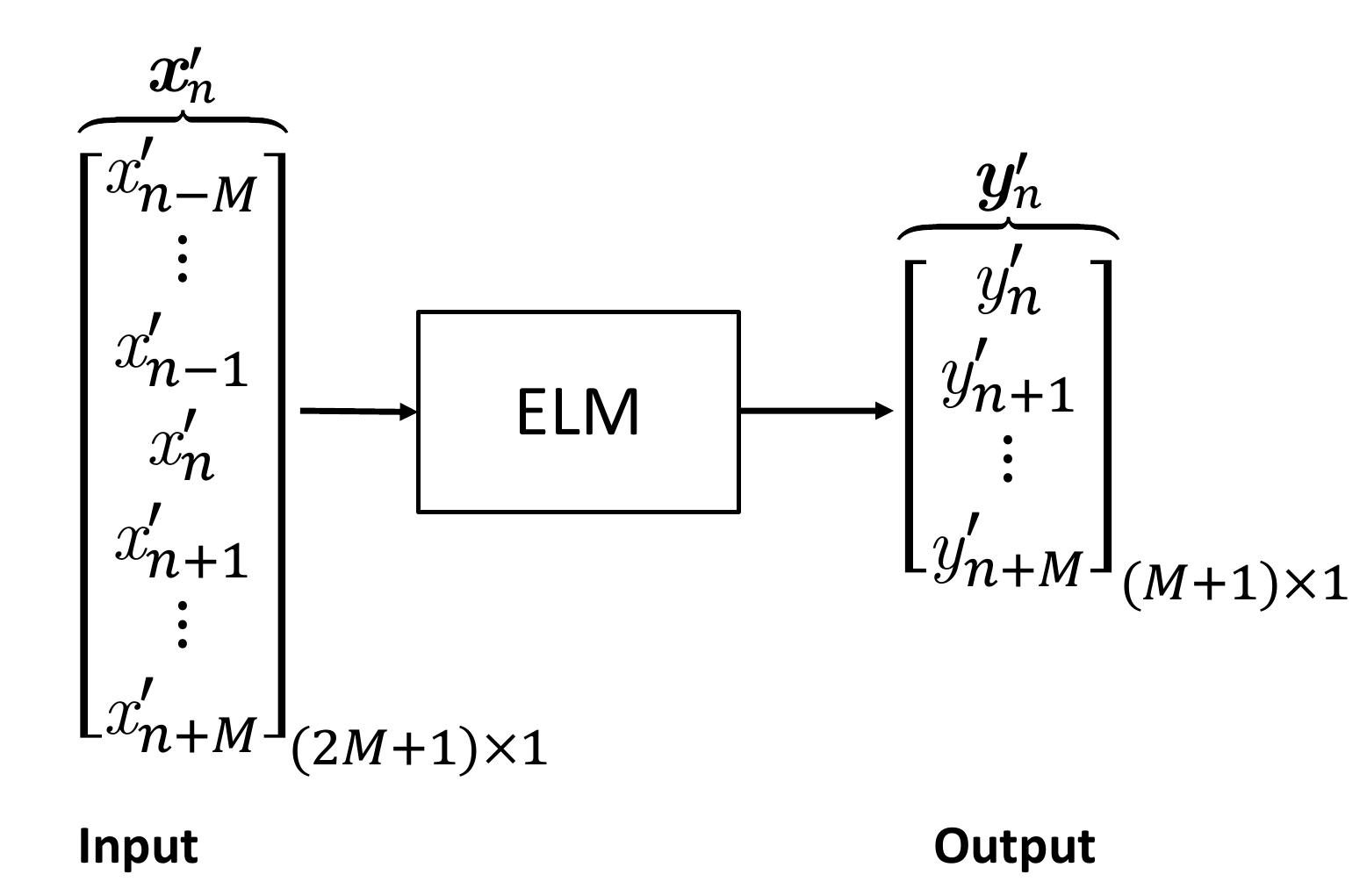}\\
			\caption{ELM training to model the LED nonlinearity and memory effects.}\label{TELM}
		\end{center}
	\end{figure}

As we are using sliding window approach, given an input vector $\bm{x}_n=[x_{n-M},\ldots,x_{n-1},x_n,x_{n+1},\ldots,x_{n+M}]^T$, we use ELM to predict the corresponding output of the LED $\bm{y}_n=[y_n,y_{n+1},\ldots,y_{n+M}]^T$. Therefore, the dimensions of ELM input and output are $2M+1$ and $M+1$, respectively, as shown in Fig.~\ref{TELM}. We can use the training sequence $\{x^{\prime}_n\}$ and the corresponding received signals $\{y^{\prime}_n\}$ to train the ELM, and the output weights $\bm{\beta}^\ast$ can be obtained based on~(\ref{beta}). The trained ELM models the LED nonlinearity and memory effects, i.e., with the input vector $\bm{x}_n=[x_{n-M},\ldots,x_{n-1},x_n,x_{n+1},\ldots,x_{n+M}]^T$, the output of the ELM is approximately $\bm{a}(x_n)+{\mathbf{{i}}}_{n}$, where $\bm{a}(x_n)$ and ${\mathbf{{i}}}_{n}$ are defined in~(\ref{MI}). This enables us to estimate $\bm{a}(\alpha_i)$, $\overline{\mathbf{{i}}}_{n}$ and $\bm{V}$ to evaluate the likelihood $p(\bm{y}_n\lvert x_n )$ in~(\ref{pyx}), which is detailed in the following.
	\subsubsection{Estimate $\bm{a}(\alpha_i)$ based on the trained ELM}	
	Note that the input of the trained ELM is a vector $\bm{x}_n$, and the interference $\mathbf{i}_n$ is determined by $\{x_j,j\neq n\}$. Then $\bm{a}(\alpha_i)$ can be estimated by setting $x_n=\alpha_i$ and $x_j=0, j\neq n$. Hence, with a special input vector $\bm{x}_n=[0,\ldots,0,\alpha_i,0,\ldots,0]^T$, the output of the ELM can be used as an estimate of $\bm{a}(\alpha_i)$, which is illustrated in Fig.~\ref{subfig:a}.
	\subsubsection{Estimate $\overline{\mathbf{{i}}}_{n}$ based on the trained ELM}
	We model the interference $\mathbf{i}_n$ as 
	\begin{equation}\label{inm}
	\mathbf{i}_n = \hat{\mathbf{i}}_n+\bm{w}_{\mathbf{i}_n},
	\end{equation}
	where $\hat{\mathbf{i}}_n$ is an estimate of ${\mathbf{i}}_n$, and $\bm{w}_{\mathbf{i}_n}$ denotes a zero mean noise. Hence, $\Bar{\mathbf{i}}_n=\hat{\mathbf{i}}_n$. Next, we use the trained ELM to find the estimate of ${\mathbf{i}}_n$. As we know the interference ${\mathbf{i}}_n$ is determined by $x_j, j\neq n$ in the vector $\bm{x}_n$, in order to get $\hat{\mathbf{i}}_n$, we need the estimate of $x_j, j\neq n$, which is denoted by $\hat{x}_j$. As we know in a turbo system, $\hat{x}_j$ can be calculated based on the feedback from the decoder, which is the mean of $x_j$~\cite{539767}, i.e.,
	\begin{equation}\label{llr}
	\hat{{x}}_{j} = \sum \limits^{2^P}_{i=1}\alpha_i p(x_j=\alpha_i),
	\end{equation}
	where $p(x_j=\alpha_i)$ can be calculated based on extrinsic LLR obtained at each iteration. Therefore, $\hat{\mathbf{i}}_n$ can be obtained based on the trained ELM with a special input 
	$\bm{x}_n=[\hat{{x}}_{n-M},\cdots,\hat{{x}}_{n-1},0,\hat{{x}}_{n+1},\cdots,\hat{{x}}_{n+M}]^T$, where we set $x_n=0$. This is shown in Fig.~\ref{subfig:ilight}.
	\subsubsection{Estimate $\bm{V}$ based on the trained ELM} Recall that $\bm{V}$ defined in (16) is the covariance matrix of interference ${\mathbf{{i}}}_{n}$ plus noise $\bm{w}_n$. It can be estimated as
	\begin{equation}
	\hat{\bm{V}} =\frac{1}{N}\sum\limits_{n=1}^N(\bm{y}_n-\bm{a}(x_n)-\overline{\mathbf{{i}}}_{n})(\bm{y}_n-\bm{a}(x_n)-\overline{\mathbf{{i}}}_{n})^T.
	\end{equation}
	\begin{figure}
		\centering
		
		\subfloat[Estimate $\bm{a}(\alpha_i)$ based on the trained ELM.]{
			\label{subfig:a}
			\includegraphics[width=0.4\textwidth]{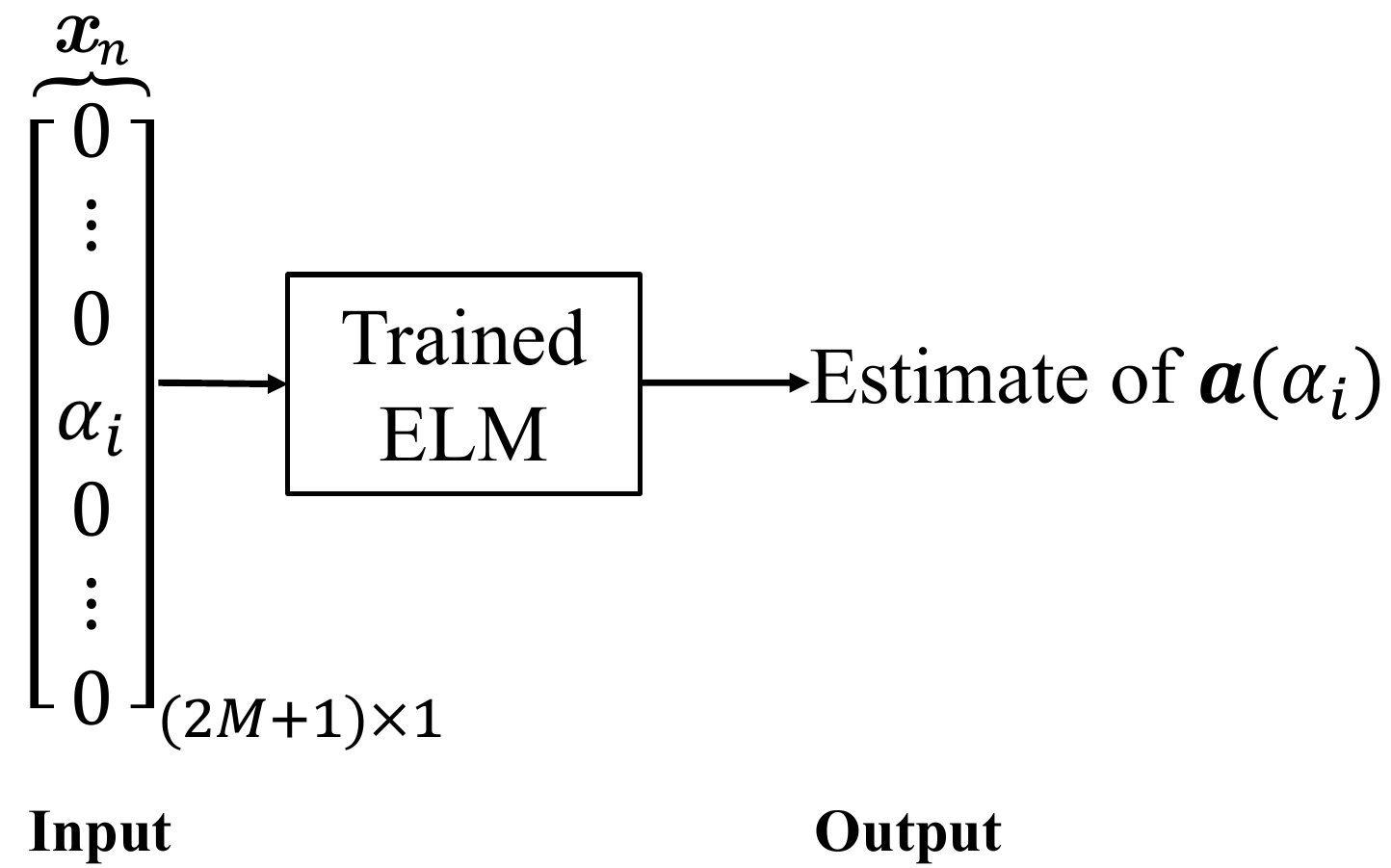} }
		
		\subfloat[Estimate $\overline{\mathbf{{i}}}_n$ based on the trained ELM.]{
			\label{subfig:ilight}
			\includegraphics[width=0.4\textwidth]{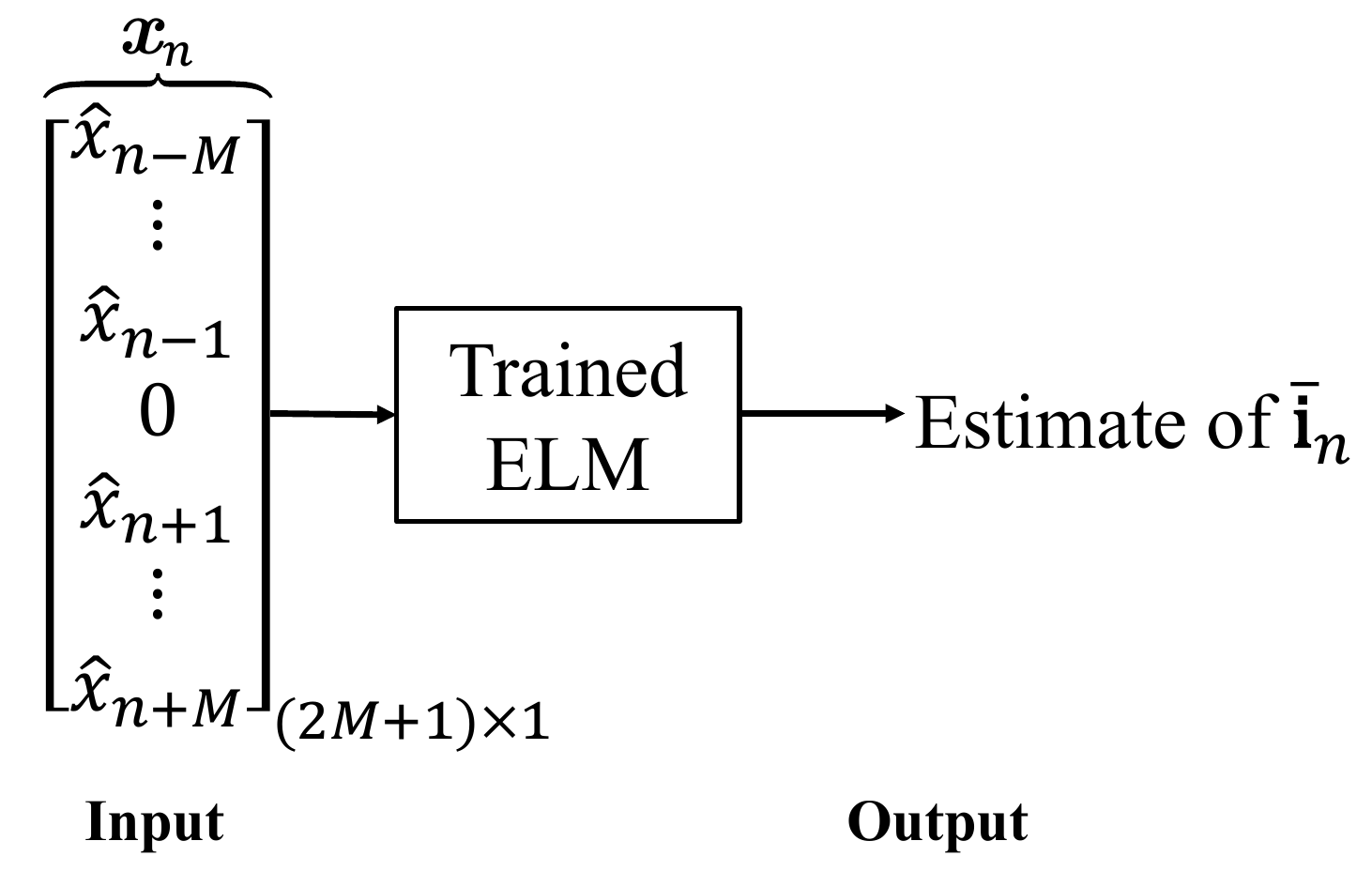}} 
		
		\subfloat[Estimate $\bm{a}({x}_{n})$ based on the trained ELM.]{
			\label{subfig:xb}
			\includegraphics[width=0.4\textwidth]{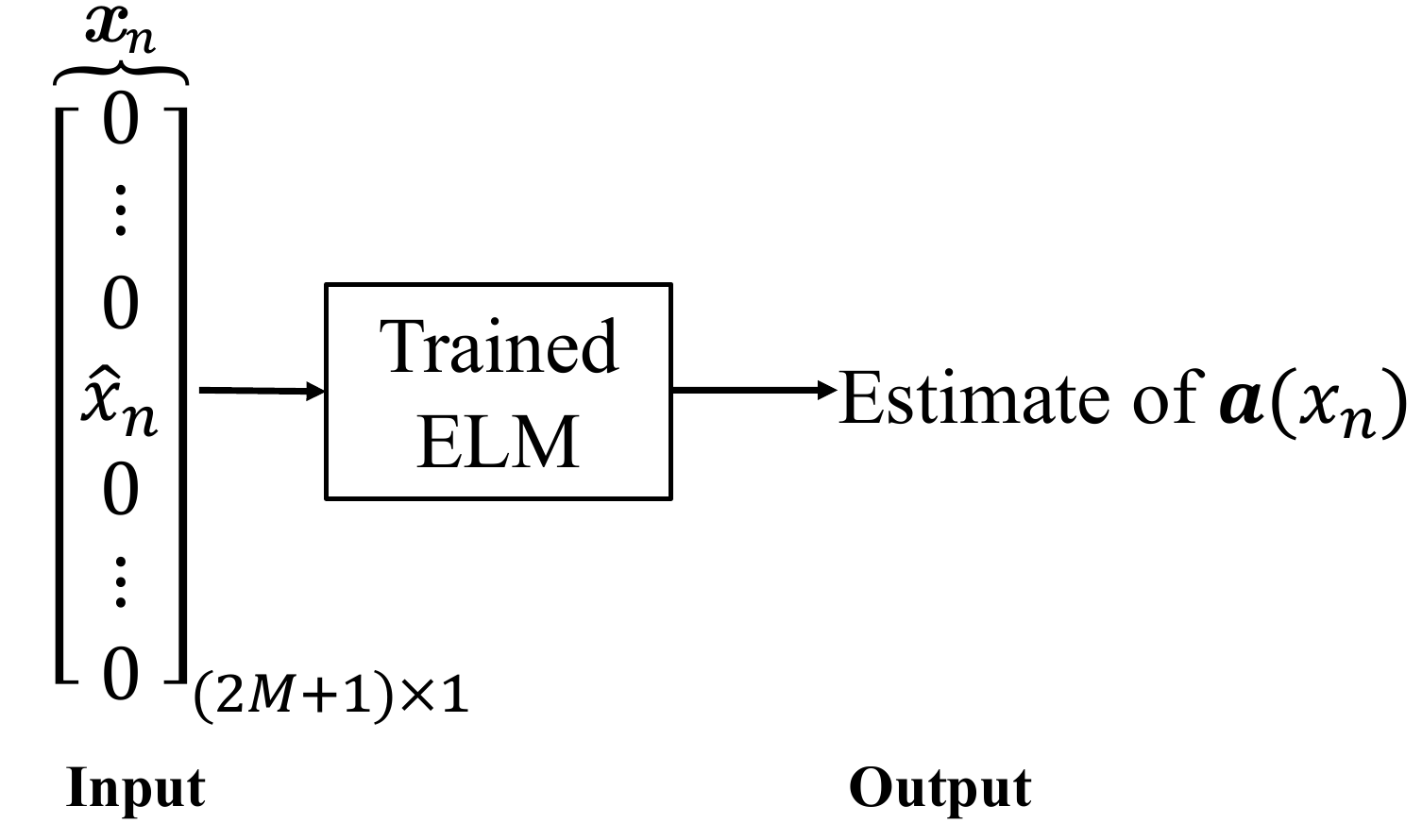}} 
		\caption{Input and output of trained ELM for estimating the required variables in~(\ref{pyx}).} 
		\label{kohler}
	\end{figure}
	However $\bm{a}(x_n)$ is unavailable, because $x_n$ is unknown. We note that the estimate of $x_n$, i.e.,  $\hat{x}_{n}$, can be calculated  from the feedback from the decoder, so a practical estimator for $\hat{\bm{V}}$ is 
	\begin{equation}
	\hat{\bm{V}} =\frac{1}{N}\sum\limits_{n=1}^N(\bm{y}_n-\bm{a}(\hat{x}_{n})-\overline{\mathbf{{i}}}_{n})(\bm{y}_n-\bm{a}(\hat{x}_{n})-\overline{\mathbf{{i}}}_{n})^T.
	\end{equation}
	In addition, to reduce the complexity in computing~(\ref{pyx}), we further approximate $\hat{\bm{V}}$ as a diagonal matrix by ignoring its off-diagonal elements. Hence, the matrix inverse $\hat{\bm{V}}^{-1}$ is trivial and the evaluation of~(\ref{pyx}) actually only involves scalar operations, thus leading to low complexity. Noting that $\Bar{\mathbf{i}}_n$ is already obtained in the above, we only need to calculate $\bm{a}(\hat{x}_n)$. This can be easily achieved by using the trained ELM with the input vector $\bm{x}_n=[0,\ldots,0,\hat{{x}}_{n},0,\ldots,0]^T$, which is illustrated in Fig.~\ref{subfig:xb}.
	\section{Data-aided ELM Based Iterative Receiver}
	We need sufficient training samples to train the ELM so that it can accurately model the LED nonlinearity and memory effects. By taking advantage of the iterative receiver, the estimated data based on the feedback from the decoder can be exploited as virtual training sequence for ELM training, i.e., ELM can be trained with both training sequence and data. This can significantly reduce the length of training sequence, thereby improving transmission efficiency. 
	Different from Section~\ref{ss}, where only the training sequence (it is exactly known) is used, we have to consider the uncertainty of the virtual training sequence because they are estimated and refined through the iterative process at the receiver. In this case, the estimation of ELM output weight $\bm{\beta}$ is based on the following model,
	\begin{figure}
		\begin{center}
			\includegraphics[width=3.5in]{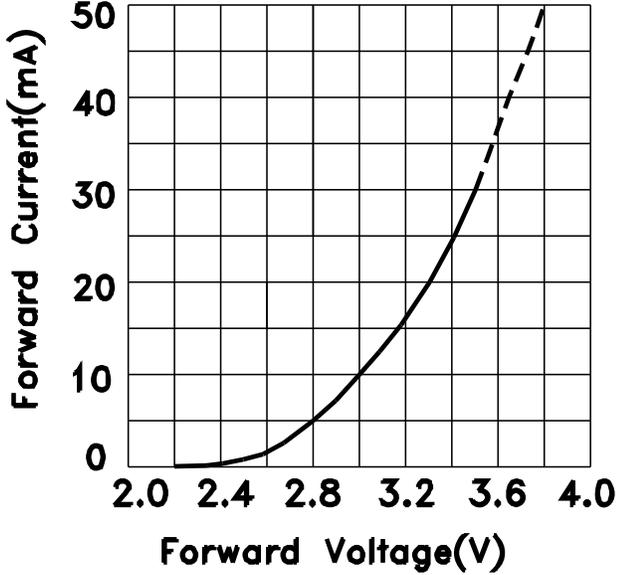}\\
			\caption{I-V response of a commercial LED (Kingbright blue T-1 3/4 (5mm) LED)~\cite{IEEEexample:IEEEwebsite}.}\label{D2}
		\end{center}
	\end{figure}
	\begin{equation}\label{mod}
	\underbrace{ (\bm{H}_t+\Delta \bm{H})}_{\bm{H}}\bm{\beta} = \underbrace{\bm{Y}_t+\Delta \bm{Y}}_{\bm{Y}},
	\end{equation}
	where $\Delta \bm{H}$ accounts for the hidden perturbation matrix due to the uncertainty of the virtual training sequence, $\Delta \bm{Y}$ denotes unknown error between the observed signal $\bm{Y}$ and the true signal $\bm{\bm{Y}_t}$.
	The total least squares (TLS) can be used for fitting such a model to data that minimizes errors in both the transmitted sequence and observed sequence~\cite{717073}. The problem is now to find the smallest perturbations $[\Delta \bm{H}~ \Delta \bm{Y}]$ to the measured signals that satisfy the model in~(\ref{mod}), i.e.,
	\begin{equation}\label{TLSS}
	\text{min}_{\bm{\beta}}\left\lVert [\Delta \bm{H}~ \Delta \bm{Y}] \right\rVert^2,~\text{such that}~(\bm{H}_t+\Delta \bm{H})\bm{\beta} = \bm{Y}_t+\Delta \bm{Y}.
	\end{equation}
	Singular value decomposition (SVD) can be used to find the unique solution to the TLS problem in~(\ref{TLSS})~\cite{717073, MARKOVSKY20072283}. Take the SVD of the matrix $[\bm{H}~\bm{Y}]$, i.e.,
	\begin{equation}
	[\bm{H}~\bm{Y}]=\bm{U\Sigma\Theta^T}. 
	\end{equation}
	Define the partitioning of matrix and $\bm{\Theta}$ as
	\begin{equation}
	\bm{\Theta}=\begin{bmatrix}
	{[\bm{\theta}_{11}]}_{L\times L}&{[\bm{\theta}_{12}]}_{L\times (M+1)} \\
	{[\bm{\theta}_{21}]}_{(M+1)\times L}&{[\bm{\theta}_{22}]}_{(M+1)\times (M+1)}\\
	\end{bmatrix}.
	\end{equation}
	The TLS solution exists if and only if $\bm{\theta}_{22}$ is non-singular, and the output weights $\bm{\beta}^{\ast}$ are given by 
	\begin{equation}
	\bm{\beta}^{\ast}=-\bm{\theta}_{12}\bm{\theta}_{22}^{-1}.
	\end{equation}

	Note that the proposed data-aided ELM based iterative receiver consists of two phases: an initialization phase and a data-aided phase. In the initialization phase, the proposed non-data aided receiver in the Subsection~\ref{ss} is used to estimate the nonlinearity merely using the training sequence due to the lack of feedback from the decoder. 
	Following the initialization phase, the data-aided training phase commences and both the training sequence and the estimate of the data sequence calculated based on the feedback from the decoder are used to re-train the ELM to improve the modelling accuracy for the LED nonlinearity and memory effects.
	\section{Experimental Results}
	\begin{figure}
		\begin{center}
			\includegraphics[width=3.7in]{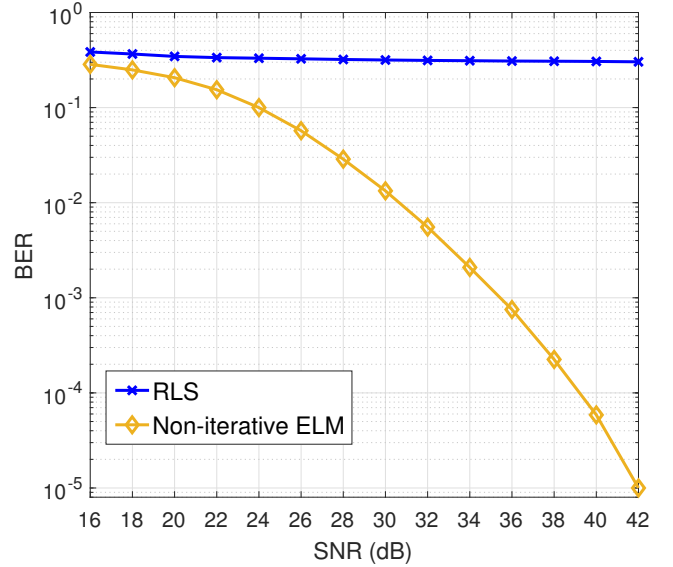}\\
			\caption{BER performance of non-iterative receivers.}\label{no1}
		\end{center}
	\end{figure}
	In this section, we present experimental results to demonstrate the effectiveness of the proposed non-iterative and iterative ELM based receivers in handling LED nonlinearity and memory effects. A convolutional code with generator [171, 133] is used for the coding scheme, and the APP decoder is implemented using the BCJR algorithm~\cite{1055186}. A high order modulation 8-PAM is used.
	We consider a commercial LED (Kingbright blue T-1 3/4 (5mm) LED) whose I-V curve (extracted from the datasheet~\cite{IEEEexample:IEEEwebsite}) is shown in Fig~\ref{D2}. The Hammerstein model is employed to include both LED nonlinearity and memory effects in the simulations~\cite{690370211} (Noting that other models can also be used to generate the distorted signal). The Hammerstein model is composed of a nonlinear block followed by a linear time-invariant (LTI) block, which is expressed as 
	\begin{equation}
	z_n=\sum_{k=1}^Ka_{k}x_{n}^k+ \rho_1 \sum\limits_{k=1}^Ka_kx_{n-1}^k+ \rho_2 \sum\limits_{k=1}^Ka_kx_{n-2}^k,
	\end{equation}
	where $\rho_1=0.15$ and $\rho_2=0.05$. AWGN is added at the receiver side. The signal-to-noise ratio (SNR) is defined as $\text{E}(z_n^2)/\sigma_n^2$, where $\sigma_n^2$ is the variance of the noise. We use $sin$ as the activation function in the hidden layer for ELM and its input weights and biases are randomly distributed from $[-1,1]$.
	\begin{figure}
		\begin{center}
			\includegraphics[width=3.7in]{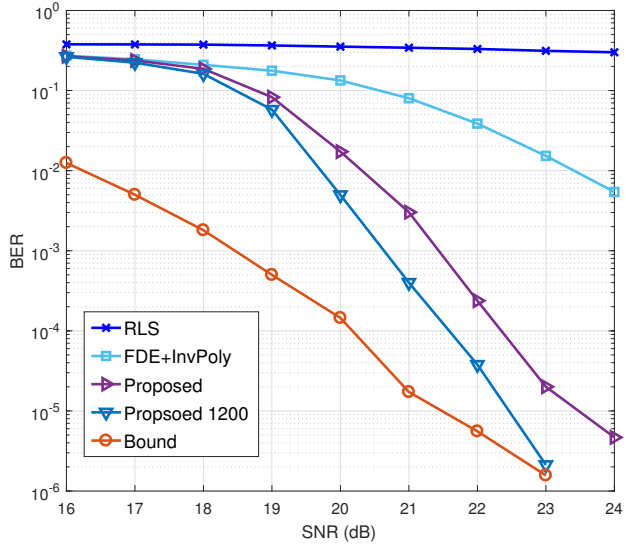}\\
			\caption{BER performance of various receivers.}\label{R1}
		\end{center}
	\end{figure}
	
	We first examine the performance of non-iterative receivers in Section~\ref{pol} with a training length of 800. The conventional RLS based post-distorter~\cite{qian2014adaptive} is implemented with the order $\Tilde{K}=7$ and the memory length $\Tilde{M} =4$. The ELM has 100 hidden nodes in the ELM based non-iterative receiver. The BER performance is shown in Fig.~\ref{no1}. We can see that the RLS based post-distorter simply does not work. In contrast, the non-iterative ELM based receiver works very well. The conventional polynomial based method can easily suffer from numerical instability, thereby leading to poor performance. However, the neural network-based method does not have such issue. From Fig.~\ref{no1}, we can see that neural network-based method is much more effective in combating the nonlinearity.
	
\begin{figure}
		\begin{center}
			\includegraphics[width=3.7in]{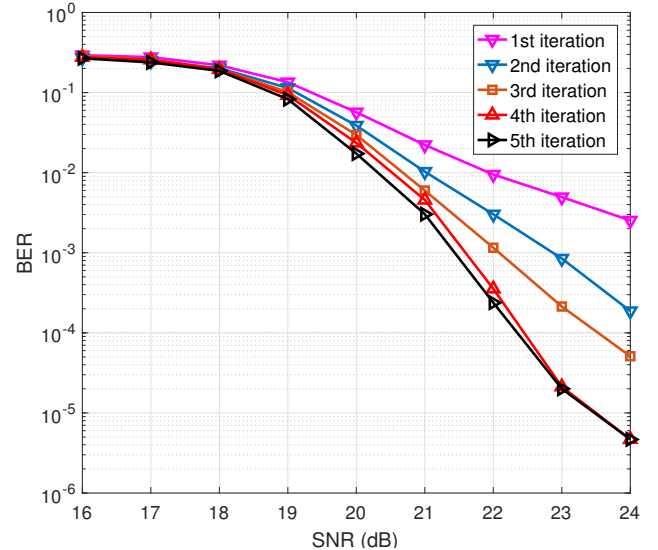}\\
			\caption{Performance of the proposed non-data aided ELM based iterative receiver with different iterations.}\label{R2}
		\end{center}
	\end{figure}

	Figure~\ref{R1} shows the BER performance of the LED system with various receivers. The performance of the iterative receiver with the perfect knowledge of the nonlinearity and memory, is also included as a performance bound. The training sequence length is 800 except the `Proposed 1200' (which is detailed later). As shown in the Fig.~\ref{R1}, the RLS based post-distorter~\cite{qian2014adaptive} does not work properly. Fig.~\ref{R1} also shows the performance of the receiver which employs the frequency domain equalization (FDE) to compensate the memory effect and the conventional polynomial inverse technique to mitigate the nonlinear distortion of LED (denoted by `FDE+InvPoly') \cite{8388717}, where nonlinearity and memory are assumed to be perfectly known. It can be seen that the receiver does not work well. In contrast, the proposed non-data-aided ELM based iterative receiver (with 150 hidden nodes) delivers much better performance. Fig.~\ref{R2} shows the BER performance of the proposed non-data-aided ELM based iterative receiver with different iterations, which demonstrates that the converge of the iterative receiver is fast, i.e., only 4-5 iterations are needed.
	
	To demonstrate that the capability of the proposed ELM based iterative receiver can be significantly improved with longer training sequence and more hidden nodes, we increase the number of hidden nodes to 300 and the length of the training sequence is increased to 1200. Its performance is also shown in Fig.~\ref{R1}, where it is denoted by `Proposed 1200'. We can see that the performance of the proposed receiver is improved further and it can approach the performance bound closely at relatively high SNRs. However, it requires high training overhead. 
	This can be overcome by using the proposed data-aided scheme where both training sequence and data are employed to train the ELM. 
	
\begin{figure}
		\begin{center}
			\includegraphics[width=3.4in]{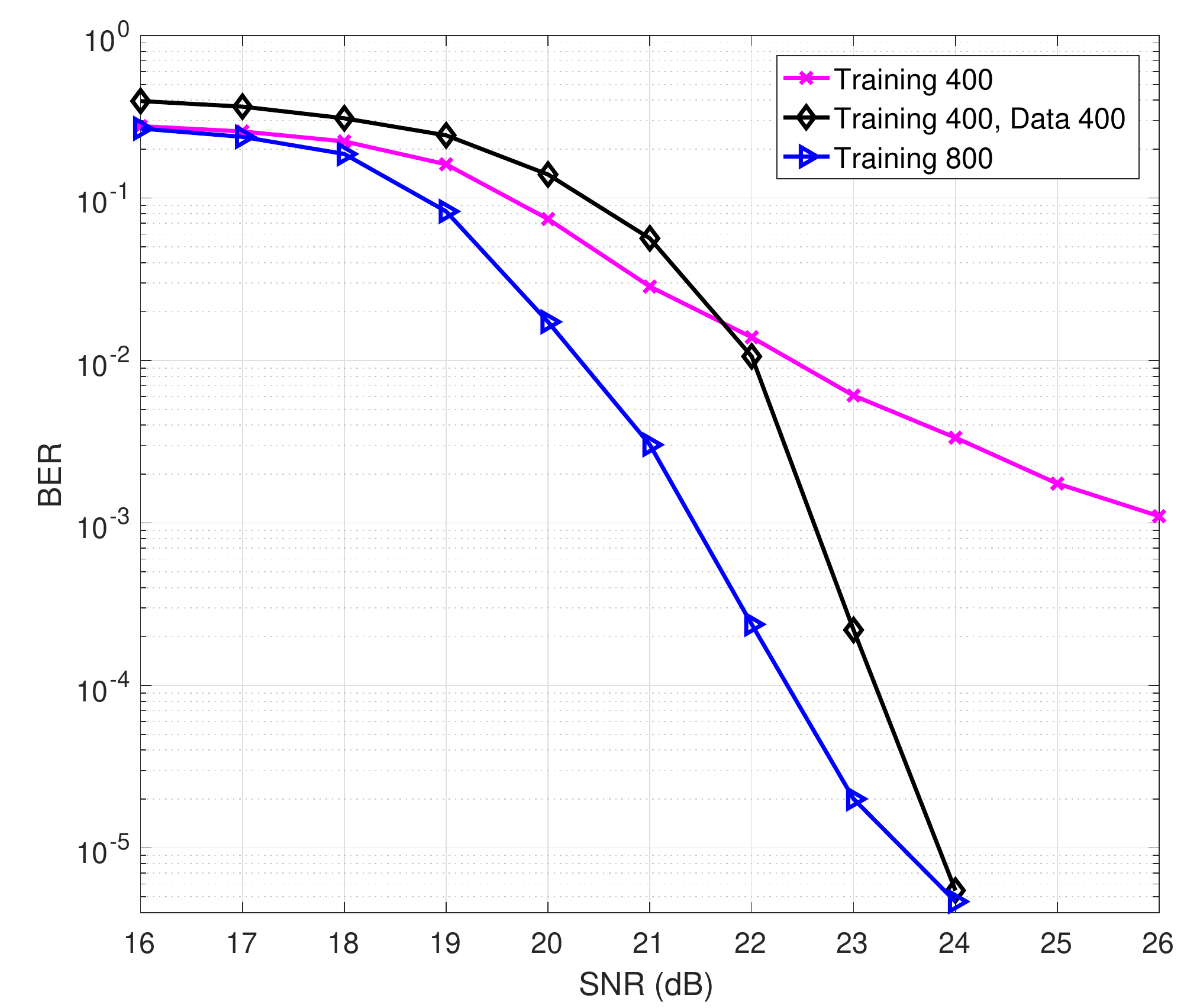}\\
			\caption{Performance comparisons of the non-data-aided and data-aided ELM based iterative receivers.}\label{R3}
		\end{center}
	\end{figure}

	Fig.~\ref{R3} shows the BER performance of both non-data-aided ELM based iterative receivers and data-aided ones with different training lengths. It can be observed that the data-aided ELM receiver with 400 training symbols and 400 data symbols outperforms the non-data-aided ELM receiver with only 400 training symbols, significantly. We can also see that the performance of the data aided receiver with 400 training and 400 data symbols can approach that of the non-data-aided ELM receiver with 800 training symbols at relatively high SNRs, i.e., training overhead is reduced by half in the data-aided ELM receiver.  
	\section{CONCLUSIONS}

	In this paper, we have proposed novel ELM based non-iterative and iterative receivers to jointly handle LED nonlinearity and memory effects. It is shown that the ELM based non-iterative receivers outperform the conventional receivers significantly, i.e., neural networks are much more efficient to deal with LED nonlinearity compared to the conventional polynomial techniques. For the iterative case, we have designed non-data aided and data-aided receivers by taking advantage of the iterative process in an iterative receiver, where data can serve as a virtual training sequence. It has been shown that a huge performance gain can be achieved by iterative receivers compared to non-iterative receivers, and the data-aided receiver can reduce considerable training overhead compared to the non-data aided receiver. It is worth mentioning that this work  can also be extended to RF communications.

	
	
	\bibliographystyle{IEEEtran}
	\bibliography{bare_jrnl}
\end{document}